\documentclass[
 reprint, superscriptaddress,
 amsmath,amssymb,
 aps,
prl,
]{revtex4-1}

\usepackage{graphicx}
\usepackage{dcolumn}
\usepackage{bm}

\usepackage{color}
\usepackage{ulem}

\newcommand{\MM}{\mathbf{M}}
\newcommand{\ud}{\mathrm{d}}

\begin{document}

\title{Tuning magnetic confinement of spin-triplet superconductivity}



\author{Wen-Chen Lin}
\thanks{These authors contributed equally.}
\affiliation{Maryland Quantum Materials Center, Department of Physics, University of Maryland, College Park, MD 20742, USA}

\author{Daniel J. Campbell} 
\thanks{These authors contributed equally.}
\affiliation{Maryland Quantum Materials Center, Department of Physics, University of Maryland, College Park, MD 20742, USA}

\author{Sheng Ran}
\affiliation{Maryland Quantum Materials Center, Department of Physics, University of Maryland, College Park, MD 20742, USA}
\affiliation{NIST Center for Neutron Research, National Institute of Standards and Technology, Gaithersburg, MD 20899, USA}
\affiliation{Department of Materials Science and Engineering, University of Maryland, College Park, MD 20742, USA}

\author{I-Lin Liu}
\affiliation{Maryland Quantum Materials Center, Department of Physics, University of Maryland, College Park, MD 20742, USA}
\affiliation{NIST Center for Neutron Research, National Institute of Standards and Technology, Gaithersburg, MD 20899, USA}
\affiliation{Department of Materials Science and Engineering, University of Maryland, College Park, MD 20742, USA}

\author{Hyunsoo Kim}
\affiliation{Maryland Quantum Materials Center, Department of Physics, University of Maryland, College Park, MD 20742, USA}

\author{Andriy H. Nevidomskyy}
\affiliation{Department of Physics and Astronomy, Rice University, Houston, Texas 77005, USA}
\affiliation{Rice Center for Quantum Materials, Rice University, Houston, Texas 77005, USA}

\author{David Graf}
\affiliation{National High Magnetic Field Laboratory, Florida State University, Tallahassee, FL 32313, USA}

\author{Nicholas P. Butch}
\affiliation{Maryland Quantum Materials Center, Department of Physics, University of Maryland, College Park, MD 20742, USA}
\affiliation{NIST Center for Neutron Research, National Institute of Standards and Technology, Gaithersburg, MD 20899, USA}

\author{Johnpierre Paglione}
\email{paglione@umd.edu}
\affiliation{Maryland Quantum Materials Center, Department of Physics, University of Maryland, College Park, MD 20742, USA}
\affiliation{Department of Materials Science and Engineering, University of Maryland, College Park, MD 20742, USA}
\affiliation{Canadian Institute for Advanced Research, Toronto, Ontario M5G 1Z8, Canada}

\date{\today}

\begin{abstract}
Electrical magnetoresistance and tunnel diode oscillator measurements were performed under external magnetic fields up to 41~T applied along the crystallographic $b$-axis (hard axis) of UTe$_2$ as a function of temperature and applied pressures up to 18.8~kbar. In this work, we track the field-induced first-order transition between superconducting and magnetic field-polarized phases as a function of applied pressure, showing a suppression of the transition with increasing pressure until the demise of superconductivity near 16~kbar and the appearance of a pressure-induced ferromagnetic-like ground state that is distinct from the field-polarized phase and stable at zero field. 
Together with evidence for the evolution of a second superconducting phase and its upper critical field with pressure, we examine the confinement of superconductivity by two orthogonal magnetic phases and the implications for understanding the boundaries of triplet superconductivity.


\end{abstract}

\maketitle


Previous work on uranium-based compounds such as UGe$_2$, URhGe and UCoGe has unearthed a rich interplay between superconductivity and ferromagnetism in this family of materials~\cite{aoki2019review}, with suggestions that ferromagnetic spin fluctuations can act to enhance pairing~\cite{Mineev2015Reentrant}. 
The recent discovery of superconductivity in UTe$_2$ has drawn strong attention owing to a fascinating list of properties -- including absence of magnetic order at ambient pressure~\cite{sundar2019coexistence}, Kondo correlations and extremely high upper critical fields~\cite{ran2019nearly} -- that have led to proposals of spin-triplet pairing~\cite{ran2019nearly,aoki2019unconventional,metz2019point,nakamine2019superconducting}, and a chiral order parameter~\cite{bae2019anomalous,jiao2019microscopic}. 
In addition, at least two forms of re-entrant superconductivity have been observed in high magnetic fields, including one that extends the low-field superconducting phase upon precise field alignment along the crystallographic $b$-axis~\cite{Knebel2019Field}, and an extreme high-field phase that onsets in pulsed magnetic fields above the paramagnetic normal state at angles tilted away from the $b$-axis~\cite{ran2019extreme}.

Applied pressure has also been shown to greatly increase the superconducting critical temperature $T_c$ in UTe$_2$~\cite{ran2019enhanced,braithwaite2019multiple}, from 1.6~K to nearly double that value near 10 kbar, and to induce a second superconducting phase above a few kbar~\cite{braithwaite2019multiple}. Upon further pressure increase, evidence of a suppression of the Kondo energy scale leads to an abrupt disappearance of superconductivity and a transition to a ferromagnetic phase~\cite{ran2019enhanced}.
Together with the ambient pressure magnetic field-induced phenomena~\cite{ran2019extreme,knafo2019magnetic,Knebel2019Field,Miyake2019Metamagnetic},
the axes of magnetic field, temperature and pressure provide for a very rich and interesting phase space in this system. 
One of the key questions is in regard to the field-polarized (FP) phase that appears to truncate superconductivity at 34.5~T under proper $b$-axis field alignment~\cite{ran2019extreme,Knebel2019Field}, in particular regarding the nature of the coupling of the two phases and whether superconductivity could persist to even higher fields in the absence of the competing FP phase. The relation between the FP phase and the pressure-induced magnetic phase, which also competes with superconductivity~\cite{ran2019extreme}, is similarly not yet fully understood.

In this work, we perform magnetoresistance (MR) and tunnel diode oscillator (TDO) measurements under both high hydrostatic pressures $P$ and high magnetic fields $H$ along the crystallographic $b$-axis to explore the ($H, T, P$) phase diagram. We find that the FP phase that interrupts superconductivity at ambient pressure is strengthened with increasing pressure, so as to suppress the transition field until there is no trace of superconductivity down to 0.4 K above 16~kbar. At higher pressures, we find evidence of a distinct magnetic phase that appears to be ferromagnetic in nature and is also bordered by the FP phase at finite fields. Together with previous observations at ambient pressure, these results suggest a spectrum of magnetic interactions in UTe$_2$ and a multi-faceted ground state sensitive to several physical tuning parameters.



\begin{figure*}[t]
\centering
\includegraphics[width=15cm, trim = 20  0  20  0, clip]{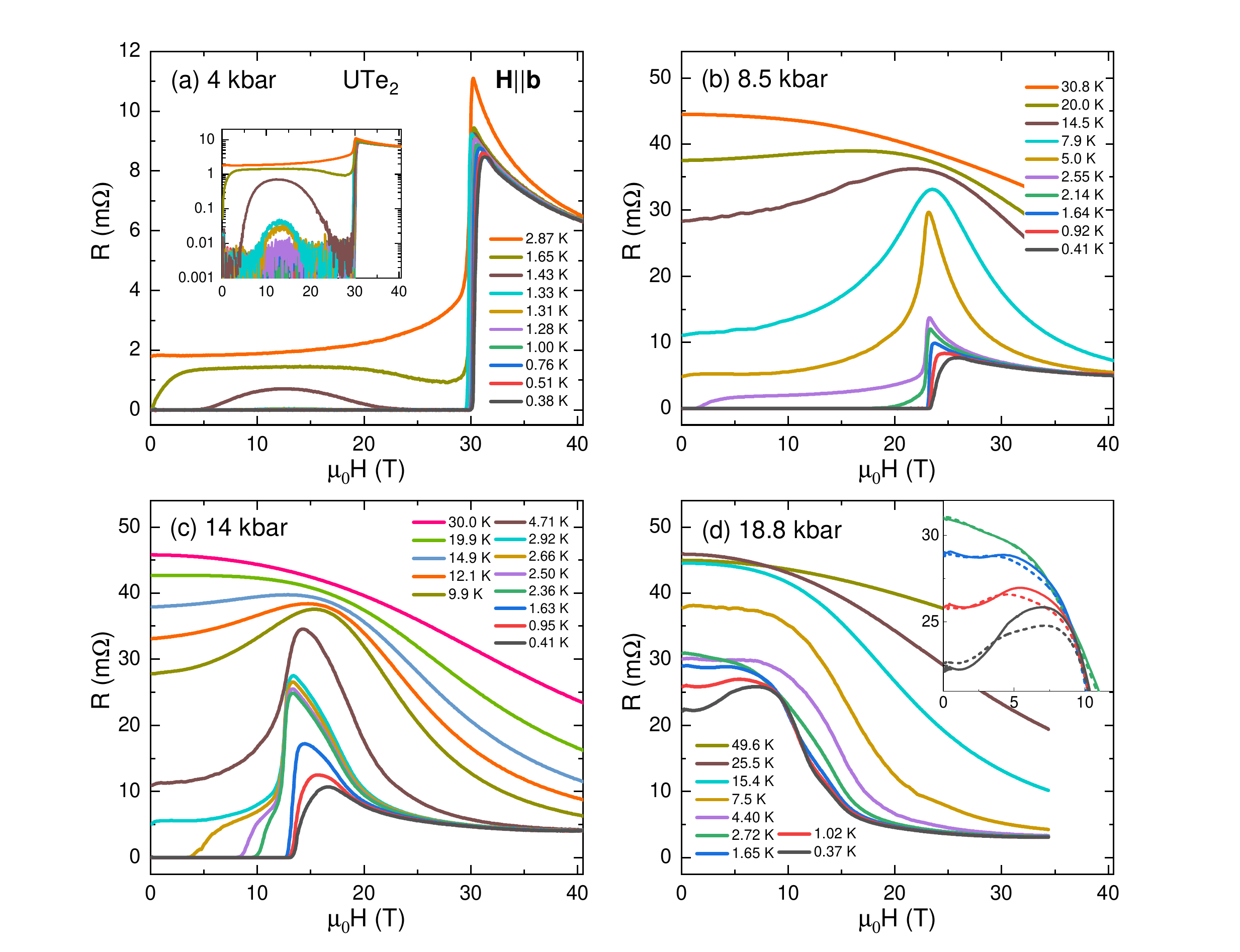}
\caption{Magnetoresistance of a UTe$_2$ single crystal with current applied along crystallographic $a$-axis and magnetic fields applied along the $b$-axis under applied pressures of (a) 4~kbar, (b) 8.5~kbar, (c) 14~kbar and (d) 18.8~kbar. Inset of (a) shows a semilog plot of magnetoresistance at 4~kbar, highlighting re-entrant superconductivity. Inset of (d) presents a zoom in the range where hysteresis is observed via distinct upsweep (solid lines) and downsweep (dashed lines) curves.}
\end{figure*}

Single crystals of UTe$_2$ were synthesized by the chemical vapor transport method as described previously~\cite{ran2019nearly}. The crystal structure of UTe$_2$ is orthorhombic and centrosymmetric, and the magnetic easy axis is the \textit{a}-axis. Experimental measurements were conducted at the DC Field Facility of the National High Magnetic Field Laboratory (NHMFL) in Tallahassee, Florida, using a 41 T resistive magnet with a helium-3 cryostat. Resistance and magnetic susceptibility measurements were performed simultaneously on two individual samples from the same batch positioned in a non-magnetic piston-cylinder pressure cell. The pressure medium was Daphne 7575 oil, and pressure was calibrated at low temperatures by measuring the fluorescence wavelength of ruby, which has a known temperature and pressure dependence~\cite{piermarini1975calibration,ragan1992calibration}.
The TDO technique uses an LC oscillator circuit biased by a tunnel diode whose resonant frequency is determined by the values of LC components, with the inductance L given by a coil that contains the sample under study; the change of its magnetic properties results in a change in resonant frequency proportional to the magnetic susceptibility of the sample.
Although not quantitative, the TDO measurement is indeed sensitive to the sample's magnetic response within the superconducting state where the sample resistance is zero~\cite{Kim2013penetration,Cho2011anisotropic, prommapan2011magnetic}.
Both the current direction for the standard four-wire resistance measurements and the probing field generated by the TDO coil are along crystallographic \textit{a}-axis (easy axis). The applied dc magnetic field was applied along the $b$-axis (hard axis) for both samples. 


The magnetic field response of electrical resistance \textit{R} at low pressures is similar to previous results at ambient pressure, which showed that the superconducting state persists up to nearly 35~T for $H\parallel b$, and re-entrant behavior can be observed near $T_c$ for slight misalignment of the field~\cite{Knebel2019Field}. As shown in Fig.~1(a), application of 4~kbar of pressure reduces the cutoff field $H^*$ to 30~T at 0.38~K ($T_c=$~1.7~K without applied field), but retains the very sharp transition to the FP state above which a negative MR ensues. Upon temperature increase, a re-entrant feature emerges below $H^*$ similar to previous reports~\cite{Knebel2019Field} but only above about 1.3~K, indicating either nearly perfect alignment along the $b$-axis or a reduced sensitivity to field angle at finite pressures. 

Upon further pressure increase, $T_c$ increases as previously shown~\cite{ran2019enhanced,braithwaite2019multiple}, up to 2.6~K and 2.8~K at 8.5~kbar and 14~kbar, respectively. However, $H^*$ is continuously reduced through this range and changes in character. As shown in Fig.~1(b) and (c), at higher pressures $H^*$ and $H_{c2}$ dissociate, beginning as a single sudden rise with a broadened peak (denoted $H_p$) in resistance at 0.4~K that becomes better-defined upon increasing from lowest temperature, before separating into two distinct transitions at higher temperatures. Interestingly, the transition is the sharpest when the $H_{c2}$ transition separates from $H^*$ and moves down in field. Further, the coupled transitions slightly decrease in field until about 2~K, above which the resistive $H_{c2}$ continues to decrease while $H^*$ stalls (e.g. at about 12~T for 14~kbar) until washing out above approximately 20~K.
This indicates a strong coupling between the two transitions that is weakened both on pressure increase and temperature increase, despite the first-order nature of the FP phase.
At 18.8~kbar, shown in Fig.~1(d), where no superconducting phase is observed down to 0.37~K, the sharp feature associated with $H^*$ is gone, and only a broad maximum in $R$ remains near 8~T. 


Figure~2 presents the frequency variation $\Delta f$ in the TDO signal, which is due to the changes in magnetic susceptibility of the sample and therefore sensitive to anomalies in the zero-resistance regime.
In addition to a sharp rise at $H^*$, which corresponds to a diamagnetic to paramagnetic transition, 
and changes in slope consistent with the re-entrant behavior mentioned above [Fig.~S3 in SI], there is another feature in the 4~kbar data within the superconducting state observable at lower fields. At temperatures below 1~K, $\Delta f$ initially increases with field before abruptly transitioning to a constant above a characteristic field $H_{c2(2)}$, and finally jumping at the $H^*$ transition. As temperature is increased, $H_{c2(2)}$ decreases in field value until it vanishes above $T_c$, tracing out an apparent phase boundary {\it within the superconducting state}. As shown in Fig.~3, the path of $H_{c2(2)}$ merges with the zero-field critical temperature of the second superconducting phase ``SC2'' discovered by ac calorimetry measurements \cite{braithwaite2019multiple}. As shown in Fig.~3(a), these data identify SC2 as having a distinct $H_{c2}(T)$ phase boundary from the higher-$T_c$ ``SC1'' phase, with a zero-temperature upper critical field of approximately 11~T at 4~kbar. Upon further pressure increase, the $H_{c2(2)}$ transition is suppressed in field, tracing out a reduced SC2 phase boundary [TDO data for 8.5~kbar in SI] that is absent by 14~kbar. In essence, it appears that the SC2 phase is suppressed more rapidly than the SC1 phase, which will provide insight into the distinction between each phase~\cite{hayes2020weyl}.

\begin{figure}[t]
\centering
\includegraphics[width=7cm, trim=0 0 0 0, clip]{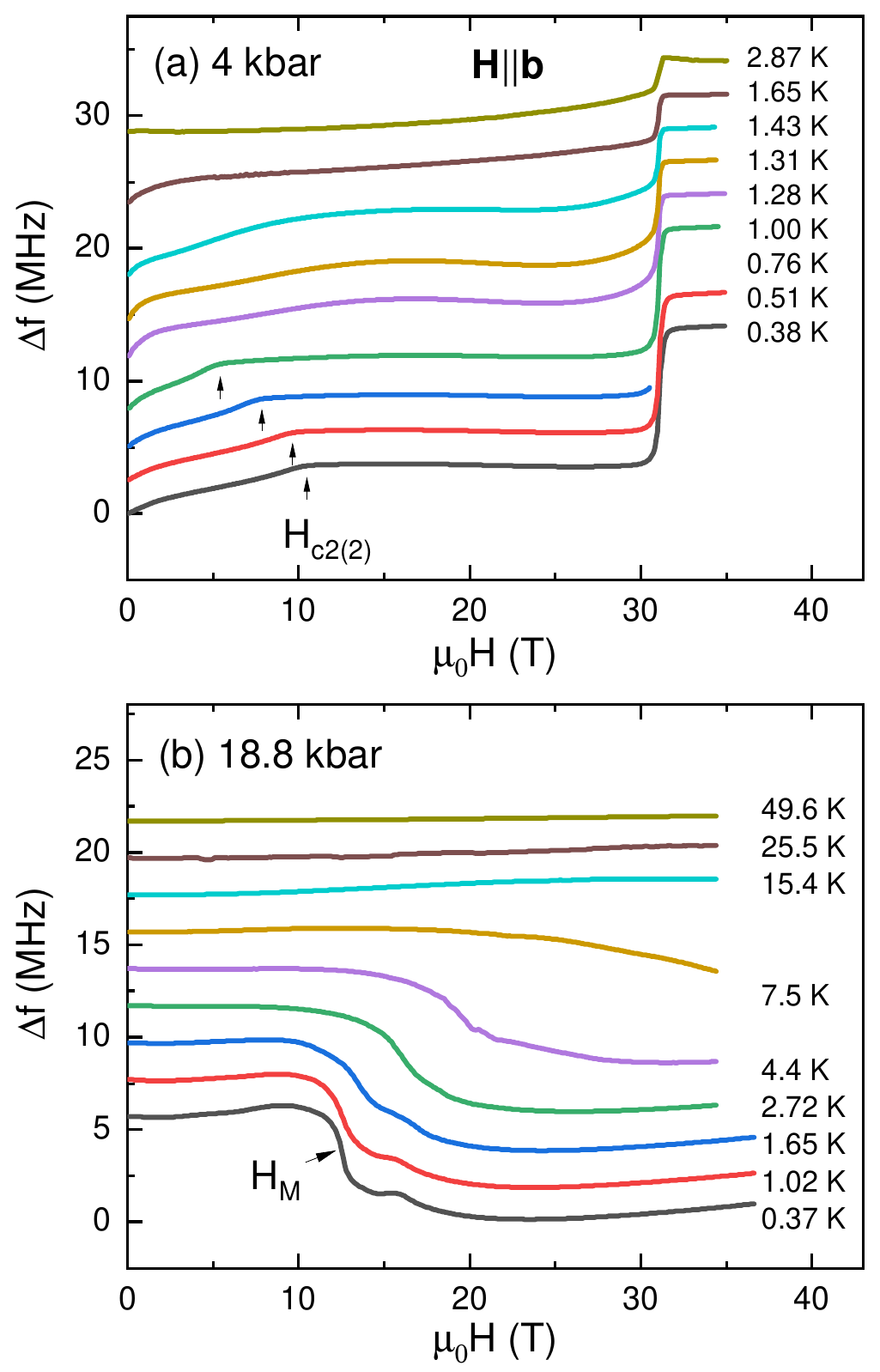}
\caption{Tunnel diode oscillator frequency variation of 
UTe$_2$ single crystal as a function of magnetic fields applied along the crystallographic $b$-axis, under applied pressures of (a) 4~kbar and (b) 18.8~kbar. Transitions involving the SC2 superconducting phase are labelled as $H_{c2(2)}$ in panel (a), and crossovers between the ferromagnetic and field-polarized phases (see text) labelled as $H_M$ in panel (b). All curves are vertically shifted for presentation.}
\end{figure}

In contrast to the abrupt increase of $\Delta f$ upon crossing $H^*$ into the FP phase at lower pressures, the TDO signal exhibits a qualitatively different response in the high pressure regime where superconductivity is completely suppressed. As shown in Fig.~2(b), at 18.8~kbar $\Delta f$ {\it decreases} at a characteristic field $H_M$ (=~12.5~T at 0.37~K), indicating a decrease of magnetic susceptibility upon entering the FP phase that is opposite to the increase observed in $\Delta f$ at lower pressures (e.g. from the normal state above $T_c$ to the FP state, in Fig.~2a). The drop at $H_M$ increases in field value and gradually flattens out as temperature increases, consistent with a ferromagnetic-like phase transition that gets washed out with magnetic field. Based on observations of hysteresis in transport (Fig.~1(d) inset) that are consistent with this picture, as well as evidence from previous pressure experiments identifying similar hysteretic behavior \cite{ran2019enhanced}, we label this phase as a ferromagnetic (FM) ground state that evolves from zero temperature and zero magnetic field, and, similar to superconductivity at lower pressures, is truncated by the FP phase and therefore distinct from that ground state.

\begin{figure*}[t]
\centering
\includegraphics[width=15cm, trim=0 10 20 0,clip]{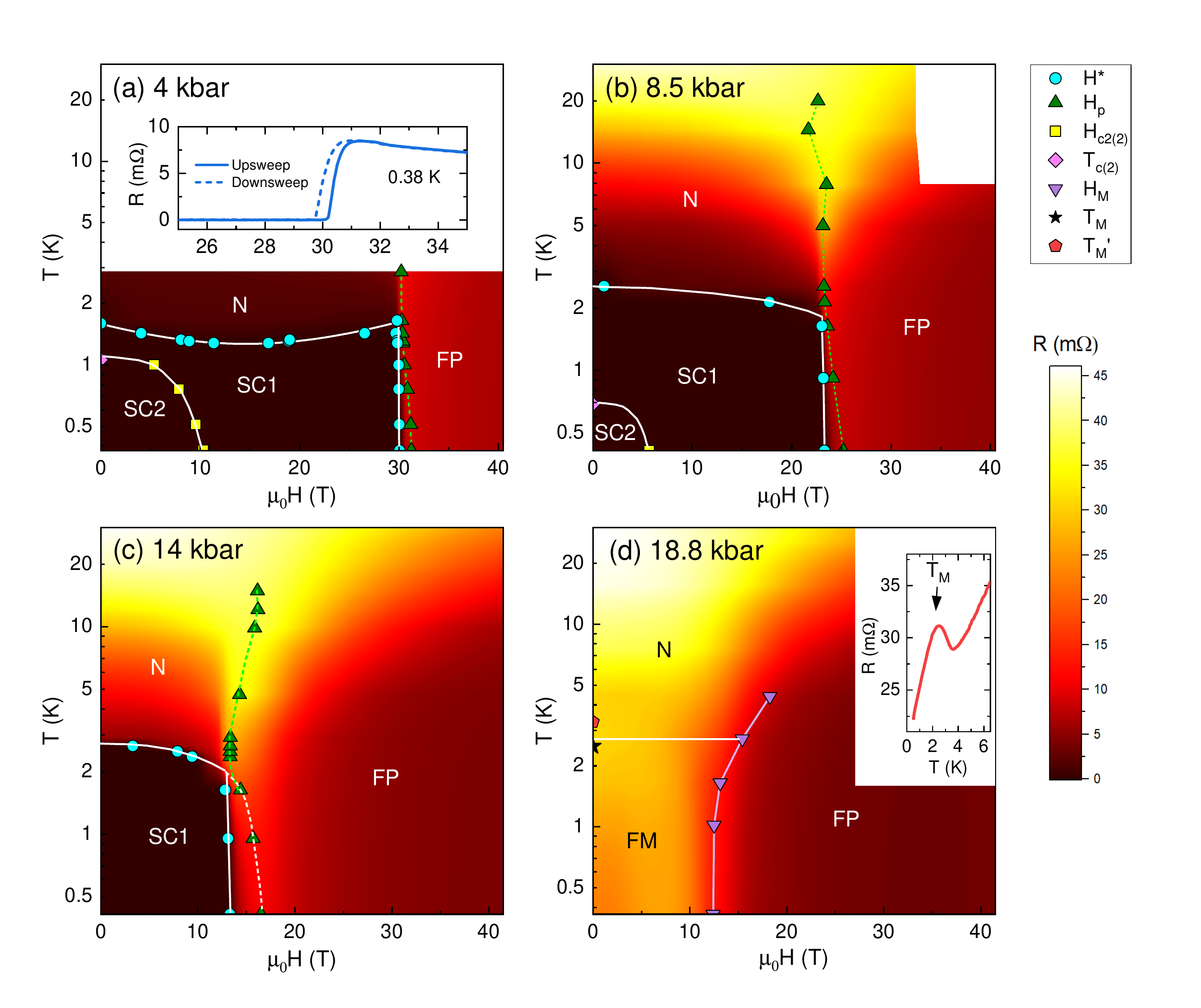}
\caption{Evolution of the magnetic field-temperature phase diagram of UTe$_2$ as a function of pressure for fields applied along the crystallographic $b$-axis, with phase boundaries of superconducting (SC1 and SC2), normal (N), field-polarized (FP) and ferromagnetic (FM) phases determined by resistance and TDO data, and concomitant variations in resistance shown by background color contours. Insert of panel (a) shows the upsweep and downsweep of magnetoresistance around the metamagnetic transition.
In panels (a)-(c), the cyan circles indicate the $T_c$ transition into the SC1 superconducting phase obtained by field sweeps which are determined by zero-resistance criteria.
Yellow squares in panels (a)-(b) indicate critical field $H_{c2(2)}$  of the superconducting phase SC2 based on TDO measurements (c.f. Fig.~2(a)), with pink diamonds indicating critical temperature $T_{c(2)}$ obtained from Ref.~\cite{braithwaite2019multiple}. 
Green triangles label the position $H_p$ of the peak in magnetoresistance in panels (a)-(c), and the purple downward triangles label the magnetic transition $H_M$ identified in TDO measurements (c.f. Fig.~2(b)) in panel (d). The black star identifies the transition $T_M$ observed in the zero-field resistance temperature dependence (panel (d) inset) while the red pentagon indicates the same transition measured in Ref.~\cite{braithwaite2019multiple}.  
}
\end{figure*}

Compiling this data, we summarize the observed features and phase boundaries in both resistance and TDO measurements in Fig.~3. We identify five phases: two superconducting phases (labeled SC1 and SC2), the normal phase (labeled N), the FP phase and the FM phase, which is only observed at 18.8~kbar.
The first three phase diagrams (4, 8.5 and 14~kbar) show a smooth growth of the FP phase with pressure and the emergence of a more conventional (i.e. rounded) $H$-$T$ boundary of the SC1 superconducting phase. In fact, the observable evolution of $H_{c2}(T)$ at 8.5 and 14~kbar indicates a putative $H_{c2}(0)$ critical point that would end within the FP phase were it not cut off by $H^*$. 
We estimate these fields to be at least 72~T and 55~T for 8.5~kbar and 14~kbar, respectively [see SI]. In this pressure range, where the putative $H_{c2}(0)$ scale becomes comparable to the FP scale $H^*$, there are clear indications of an influence on the shape of the FP transition as noted above, despite its first-order nature (c.f. hysteresis observed at base temperature shown in Fig.~3(a) inset).
Tracking the resistance peak $H_p$ to fields above $H^*$ traces a non-monotonic curve that, when below $T_c$, mimics the extension of $H_{c2}(T)$ of the SC1 phase, again suggesting an intimate correlation between the two phases. This is corroborated by the fact that at 18.8~kbar, when superconductivity is completely suppressed, the onset of the FP phase show a more conventional monotonic evolution with increasing field and temperature.


In an effort to explain the qualitative features of the phase diagram, we consider the phenomenological Ginzburg--Landau (GL) theory describing the superconducting order parameter $\mathbf{\eta}$. For simplicity we shall consider $\eta$ to be single-component, relegating to the Supplementary Materials the consideration of a multi-component order parameter proposed theoretically for UTe$_2$~\cite{nevidomskyy2020,agterberg-private}. The free energy consists of three parts: $F = F_\text{sc}[\eta] + F_m[\MM] + F_c[\eta,\MM]$, with the first term describing the superconducting order parameter in the applied field~\cite{Mineev-Samokhin}:
\begin{equation}
    F_\text{sc}[\eta] = \alpha(T)|\eta|^2 + \frac{\beta}{2}|\eta|^4 + K_{ij}(D_i \eta)^\ast(D_j \eta) + \frac{B^2}{8\pi}, \label{eq.Fsc}
\end{equation}
with $D_i = -i\nabla_i+\frac{2\pi}{\Phi_0}A_i$ denoting the covariant derivative in terms of the vector potential $\mathbf{A}$ and $\Phi_0=hc/2e$ the quantum of the magnetic flux, where $K_{ij}=\mathrm{diag}\{K_x,K_y,K_z\}$ is the effective mass tensor in the orthorhombic crystal.
The simplest way in which the superconducting order parameter couples to the field-induced microscopic magnetization $\MM$, is via the biquadratic interaction $F_c = g \MM^2|\eta|^2$, where the internal magnetic field $\mathbf{B}/\mu_0 = \MM + \mathbf{H}$. The metamagnetic transition is described by the Landau theory of magnetization with a negative quartic term ($u, v>0$):
\begin{equation}
    F_m[\MM] = \frac{\MM^2}{2\chi(P,T)} + \frac{u}{4}\MM^4 - \frac{v}{6}\MM^6 - \mathbf{H}\cdot \MM
    \label{eq.F(m)}
\end{equation}
Taking the field $\mathbf{H}||\hat{b}$, and hence $\mathbf{A}=(Hz,0,0)$, we minimize the GL free energy to obtain the \textit{linearized} gap equation of the form
\begin{equation}
    -K_z\frac{\ud^2\eta}{\ud z^2} + K_x\left(\frac{2\pi H}{\Phi_0}\right)^2 z^2 \eta + \alpha(T)\eta + g \MM^2 \eta = 0,
\end{equation}
from which one determines the $H_{c2}'$ as the lowest eignevalue of the differential operator in a standard way, similar to the problem of Landau levels for a particle in magnetic field~\cite{Tinkham}:
\begin{equation}
    H_{c2}'(T) = H_0\left[ \frac{T_c-T}{T_c} - \frac{g}{\alpha_0} M^2(H_{c2})\right], \label{eq.Hc2}
\end{equation}
where $H_0 = -T_c\left.\frac{\ud H_{c2}}{\ud T}\right|_{T_c}$ is related to the slope of $H_{c2}$ at $T_c$ in the absence of magnetization and $\alpha_0 = \frac{\hbar^2}{2m\xi_0}$ is expressed in terms of the correlation length. The upshot of Eq.~(\ref{eq.Hc2}) is that the upper critical field is reduced from its bare value $H_0(T_c-T)/T_c$ by the presence of the magnetization $M$. The latter is a function of magnetic field, $M(H)$, to be determined from Eq.~(\ref{eq.F(m)}), and while its value depends on the phenomenological coefficients of the Landau theory, 
qualitatively the metamagnetic transition results in a sudden increase of $M$ at $H^\ast$ (by $\Delta M \approx 0.6 \mu_B$ at $H^\ast=34$~T at ambient pressure~\cite{Knebel2019Field}). This then drives $H_{c2}'$ down according to Eq.~(\ref{eq.Hc2})~\cite{note_Hc2} and pins the upper critical field at the metamagnetic transition, explaining the sudden disappearance of superconductivity at the the field $H^\ast$ that marks the onset of the FP phase in Fig.~4(c).

\begin{figure}[h]
\centering
\includegraphics[width=10cm,trim=40 50 45 10,clip]{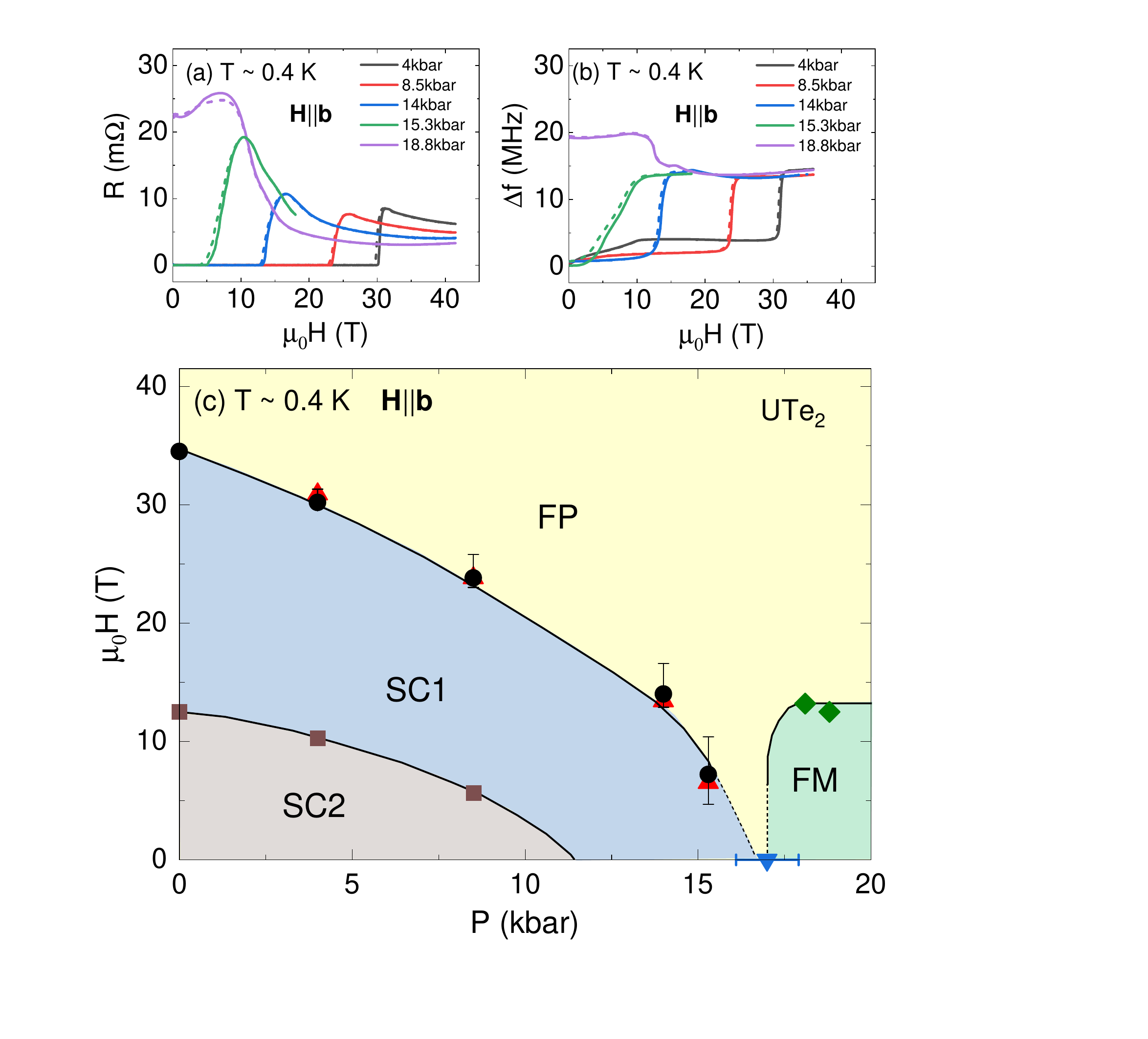}
\caption{Ground state evolution of superconducting (SC1 and SC2), field-polarized (FP) and ferromagnetic (FM) phases in UTe$_2$ as a function of applied pressure and magnetic field applied along the crystallographic $b$-axis. Panels (a) and (b) present resistance and TDO frequency variation, respectively, as functions of applied field at fixed base temperature of the measurements. Both upsweeps (solid lines) and downsweeps (dashed lines) are plotted, indicating notable hysteresis. (Note that in (b) all data are measured by a standard low-temperature-tuned TDO circuit, while 
the 15.3 kbar data was obtained using a room temperature-tuned circuit, and is therefore vertically scaled by a factor of 22 for comparison.) 
The resultant phase diagram at base temperature is presented in panel (c), where the phase boundary between SC1 and FP phases is determined by midpoints of resistance transitions (black circles, using average of upsweep and downsweep curves) and TDO transitions (red triangles), with error bars indicating width of transitions. Brown squares indicate the phase boundary of SC2 based on kinks in TDO frequency, and green diamonds indicate the transition between FM and FP phases determined from the midpoint of drops in TDO frequency response. 
Zero-pressure and zero-field data points are obtained from Refs.~\cite{ran2019extreme} and ~\cite{braithwaite2019multiple}, respectively. 
All lines are guides to the eye.}
\end{figure}

Focusing on the evolution of the ground state of UTe$_2$ with field and pressure (i.e., at our base temperature of $\sim$0.4~K), we present summary plots of the resistance and TDO data as well as the ground state field-pressure phase diagram in Fig.~4. As shown, the field boundaries of both SC1 and SC2 superconducting phases decrease monotonically with increasing pressure. However, we point out that, while the boundary of SC2 appears to be an uninterrupted upper critical field, that of SC1 is in fact the cutoff field $H^*$. It follows from Eq.~(\ref{eq.Hc2}) that this cutoff field is reduced compared to the putative $H_{c2}$, which would lie at higher fields if it were  derived from an orbital-limited model 
without taking metamagnetic transition into account.

While the $T_c$ of SC1 increases with pressure, the cutoff imposed by $H^*$ introduces difficulty in determining whether its putative $H_{c2}$ would also first increase with pressure. On the contrary, the unobstructed view of $H_{c2}$ for SC2 shows a decrease with increasing pressure that is indeed consistent with the suggested decrease of the lower $T_c$ transition observed in zero-field specific heat measurements \cite{braithwaite2019multiple}.

Between 15.3 and 18.8~kbar, the $H^*$ cutoff is completely suppressed and the FM phase onsets. While it is difficult to obtain a continuous measure of the pressure evolution through that transition, the step-like increase in the TDO frequency at a field near 12.5~T (c.f. Fig.~4(b)) measured at $P=18.8$~kbar suggests that the low-field FM phase is the true magnetic ground state of the system, separate from the FP phase. Upon closer inspection, we note that the step-like change in the TDO frequency in Figs. 2(b) and 4(b) is in fact an inflection point, suggesting that the FM and FP phases are in fact separated by a crossover, rather than a true phase transition. This is entirely natural from the Landau theory perspective, since the external magnetic field is conjugate to the FM order parameter $\MM$ in Eq.~(\ref{eq.F(m)}), and the metamagnetic crossover at field $H_M$ leads to a step-like increase in the magnetization, reflected in our TDO measurement.

This crossover boundary $H_M$ between the FM and FP phases appears much less sensitive to pressure for $P>P_c$, as evidenced by the minimal change in field value between 18.1 and 18.8~kbar. Because the experimental pressure cannot be tuned continuously, it is difficult to extract the behaviour of the crossover boundary at $P_c$. However, the previously observed discontinuity between the FM and SC1 phases as a function of pressure \cite{ran2019enhanced} suggests that the FP phase should extend down to zero field at a critical point of $P_c \sim 17$~kbar, exactly where previous zero-field work has shown an abrupt cutoff of $T_c$ and the onset of a non-superconducting phase \cite{braithwaite2019multiple}.





In summary, we have explored the pressure evolution of multiple superconducting and multiple magnetic phases of UTe$_2$ as a function of applied pressures and magnetic fields applied along the crystallographic $b$-axis, where superconductivity is known to extend to the highest fields.  The field-induced metamagnetic transition results in a field-polarized phase which cuts off superconductivity prematurely, as explained by a phenomenological  Ginzburg--Landau theory. Under increasing pressure, the superconducting phase eventually becomes completely suppressed, at the critical pressure where we observe an onset of a distinct ferromagnetic-like ground state.

\section{acknowledgments}
We thank H.-K. Wu and Y.-T. Hsu for useful discussions.
This work was performed at the National High Magnetic Field Laboratory, which is supported by the National Science Foundation Cooperative Agreement No. DMR-1644779 and the State of Florida.  
Research at the University of Maryland was supported by AFOSR grant no. FA9550-14-1-0332,  NSF grant no. DMR-1905891, the Gordon and Betty Moore Foundation’s EPiQS Initiative through grant no. GBMF9071, NIST, and the Maryland Quantum Materials Center. 
A.H.N. acknowledges support from the Robert A. Welch Foundation grant C-1818.

\bibliography{bibliography.bib}

\begin{thebibliography}{27}
\bibitem [{not()}]{note_Hc2}%
  \BibitemOpen
  \href@noop {} {}\bibinfo {note} {It should be noted that while the analysis
  of the linearized gap equation in Eq.~(4) is rather standard, it only applies
  in the vicinity of $T_c$ and cannot, strictly speaking, be used to infer the
  value of $H_{c2}(0)$ at low temperatures; at the same time, the celebrated
  Werthamer--Helfand--Hohenberg theory~\cite{WWH} establishes a proportionality
  between the value of $H_{c2}(0)$ and the slope of $H_{c2}$ and $T_c$, thus
  lending credibility to the extrapolation of Eq.~(4) to lower
  temperatures.}\BibitemShut {Stop}%
\end{thebibliography}

\end{document}